\documentclass[12pt]{article}

\usepackage{epstopdf,amsfonts,amsmath,amssymb}
\usepackage{graphicx}
\DeclareGraphicsExtensions{.eps}

\newcommand\encadremath[1]{\vbox{\hrule\hbox{\vrule\kern8pt
\vbox{\kern8pt \hbox{$\displaystyle #1$}\kern8pt}
\kern8pt\vrule}\hrule}}
\def\enca#1{\vbox{\hrule\hbox{
\vrule\kern8pt\vbox{\kern8pt \hbox{$\displaystyle #1$}
\kern8pt} \kern8pt\vrule}\hrule}}

\newcommand\framefig[1]{
\begin{figure}[bth]
\hrule\hbox{\vrule\kern8pt
\vbox{\kern8pt \vbox{
\begin{center}
{#1}
\end{center}
}\kern8pt}
\kern8pt\vrule}\hrule
\end{figure}
}

\newcommand\figureframex[3]{
\begin{figure}[bth]
\hrule\hbox{\vrule\kern8pt
\vbox{\kern8pt \vbox{
\begin{center}
{\mbox{\epsfxsize=#1.truecm\epsfbox{#2}}}
\end{center}
\caption{#3}
}\kern8pt}
\kern8pt\vrule}\hrule
\end{figure}
}
\newcommand\figureframey[3]{
\begin{figure}[bth]
\hrule\hbox{\vrule\kern8pt
\vbox{\kern8pt \vbox{
\begin{center}
{\mbox{\epsfysize=#1.truecm\epsfbox{#2}}}
\end{center}
\caption{#3}
}\kern8pt}
\kern8pt\vrule}\hrule
\end{figure}
}

\makeatletter
\@addtoreset{equation}{section}
\makeatother
\newtheorem{theorem}{Theorem}[section]

\newtheorem{remark}{Remark}[section]
\newtheorem{proposition}{Proposition}[section]
\newtheorem{lemma}{Lemma}[section]
\newtheorem{corollary}{Corollary}[section]
\newtheorem{definition}{Definition}[section]
\def\br{\begin{remark}\rm\small}
\def\er{\end{remark}}
\def\bt{\begin{theorem}}
\def\et{\end{theorem}}
\def\bd{\begin{definition}}
\def\ed{\end{definition}}
\def\bp{\begin{proposition}}
\def\ep{\end{proposition}}
\def\bl{\begin{lemma}}
\def\el{\end{lemma}}
\def\bc{\begin{corollary}}
\def\ec{\end{corollary}}
\def\beaq{\begin{eqnarray}}
\def\eeaq{\end{eqnarray}}
\newcommand{\proof}{{\noindent \bf proof:}$\quad$ }
\newcommand{\eproof}{ $\square$ }

\newcommand{\be}{\begin{equation}}
\newcommand{\ee}{\end{equation}}
\newcommand{\beq}{\begin{equation}}
\newcommand{\eeq}{\end{equation}}
\newcommand{\bea}{\begin{eqnarray}}
\newcommand{\eea}{\end{eqnarray}}

\newcommand{\ii}{{\rm i}\,}

\newcommand{\CC}{{\mathbb C}}

\newcommand{\spcurve}{{\cal S}}
\newcommand{\curve}{{\Sigma}}

\newcommand{\x}{{\rm x}}
\newcommand{\y}{{\rm y}}

\newcommand{\TR}{{\it Topological--Recursion}}

\newcommand{\td}{\tilde}
\usepackage[pdftex]{hyperref}
\hypersetup{colorlinks,urlcolor=magenta,citecolor=red,linkcolor=blue,filecolor=black}

\begin{document}

\sloppy

\pagestyle{empty}
\hfill IPhT-T19/052, Hal cea-02126908, CRM-3374, IHES-???
\addtolength{\baselineskip}{0.20\baselineskip}
\begin{center}
\vspace{26pt}
{\large \bf {Large genus behavior of topological recursion
}}
\vspace{26pt}

{\sl B.\ Eynard}${}^{123}$\hspace*{0.05cm}

\vspace{6pt}
${}^{1}$ Institut de Physique Th\'{e}orique de Saclay, UMR 3681,\\
F-91191 Gif-sur-Yvette Cedex, France.\\
${}^{2}$ CRM, Centre de recherches math\'ematiques  de Montr\'eal,\\
Universit\'e de Montr\'eal, QC, Canada,\\
${}^{3}$ IHES Bures sur Yvette.
\end{center}
\vspace{20pt}
\begin{center}
{\bf Abstract}

We show that for a rather generic set of regular spectral curves, the {\it Topological--Recursion} invariants $F_g$ grow at most like $O((\beta g)! r^{-g}) $ with some  $r>0$ and $\beta\leq 5$.

\end{center}

%

\vspace{0.5cm}

\vspace{26pt}
\pagestyle{plain}
\setcounter{page}{1}



\section{Introduction}

\TR\  \cite{EO07,Ey04,CEO05,Eynard:2008, Eyseoul,ABCD17} associates to an object called a "spectral curve" $\spcurve$, a double sequence (indexed by two non-negative integers $g,n$) of differential forms, that we shall call its "TR-invariants":
\bea
\text{TR} : \qquad  \text{Spectral curves}  & \to & \text{invariants} \cr
\spcurve & \mapsto & \{\omega_{g,n}(\spcurve)\}_{g,n} 
\eea
where  $\omega_{g,n}(\spcurve)$ is a symmetric multidifferential $n$-form, and for $n=0$, $\omega_{g,0}(\spcurve)$ is denoted $F_g(\spcurve)\in\CC $  is a complex number (a 0-form).

These invariants play an importamt role in enumerative geometry, in integrable systems, in string theory, in WKB approximation, in random matrices, ... etc, see reviews \cite{Eynard:2008, Eyseoul}.

\medskip
The main question of this article is: \textbf{how $F_g(\spcurve)$ behaves at large $g$, and more generally how $\omega_{g,n}(\spcurve)$ behaves at large $g$ ? Is the series $\sum_{g=0}^\infty \hbar^{2g-2} F_g(\spcurve)$ summable ?}
\medskip

We shall establish some bounds, under reasonable smoothness assumptions on the spectral curve $\spcurve$.
We shall find that the series
\beq
\sum_{g=0}^\infty \hbar^{2g-2} F_g(\spcurve)
\eeq
is an asymptotic series with factorially bounded coefficients, thus having a Borel transform converging in a disc.
We postpone to a following article the issue of whether this is a resurgent series and whether it can be Borel-ressumed.


\section{Bound on the growth}

\subsection{Hypothesis}

We consider a spectral curve
\beq
\mathcal S = (\curve, \x,\y,B),
\eeq
where:
\begin{itemize}
\item $\curve$ is a Riemann surface (it needs not be compact neither connected, for example it could be a union of disjoint discs, = a "local curve"),
\item $\x:\curve \to \CC P^1$ is a holomorphic function, it makes $\curve$ a ramified cover of (an open domain of) the Riemann sphere $\CC P^1$, and in particular it can have ramification points.

We shall moreover assume that $\x$ has only simple ramification points, at which the 1-form $d\x$ has only {\bf simple zeros}, and only a finite number of them, we denote the set of ramification points:
\beq
\mathfrak R = \{ a \ | \ d\x(a)=0\}.
\eeq

\item $\y$ is a meromorphic 1-form on $\curve$, that is holomorphic in a neighborhood of ramification points. We shall denote $\y=ydx$ where $y$ is thus a holomorphic function in a neighborhood of ramification points.
Remark: In the "local curve" definition of topological recursion, all what is needed is  $\y$  to be a formal series, with possibly a zero radius of convergence, here we assume something much stronger: that $\y$ is analytic in a neighborhood of every $a$. However we don't care about how $\y$ could have poles or singularities outside of these neighborhoods of $\mathfrak R$.

We shall furthermore assume that at any ramification point $a$, we have  
\beq
dy\neq 0 \quad \text{at} \quad a.
\eeq
These assumptions are generic, they indicate that near a branch point $a$, $y$ behaves like a square-root:
\beq
y(p) \sim y(a) + y'(a) \sqrt{\x(p)-\x(a)} + O(\x(p)-\x(a)) \qquad ,\quad \y=ydx.
\eeq 

\item $B$ is a meromorphic bidifferential on $\curve\times \curve$, with double pole at coinciding points, and no other poles, normalized, in any local coordinate $\zeta$ as
\beq
B(p_1,p_2) \mathop{\sim}_{p_1\to p_2} \frac{d\zeta(p_1) \otimes d\zeta(p_2)}{(\zeta(p_1)-\zeta(p_2))^2}  + \text{analytic}.
\eeq

\item 

Let us define for $p\in \curve$:
\beq
\rho(p) = \sqrt{ \prod_{a\in\mathfrak R} (\x(p)-a)}.
\eeq
For some $0<R<1$ we are going to consider the domain of $\curve$
\beq
\curve_R = \{ p \in \curve \ | \  |\rho(p)|\leq R\} .
\eeq
We assume that the radius $R$ is small enough so that $\curve_R$ is a union of disjoint discs, whose centers are the ramification points.
We make once for all a choice of squareroot in the definition of $\rho$, so that $\rho$ is analytic in each disc, and is thus a local coordinate in each disc.

\end{itemize} 

\bd
Let
\beq\label{defCKbound}
C= |\mathfrak R| \ \sup_{p,p_1\in\curve_R} \left|K(p_1,p) \frac{d\rho(p)}{d\rho(p_1)}\right| \ |\rho(p)^2-\rho(p_1)^2| \  |\rho(p)| \ 
\eeq
\beq\label{defBBbound}
B=\sup_{p,p_1\in\curve_R} \left| \frac{B(p_1,p)}{d\rho(p) d\rho(p_1)}\right| \ |\rho(p)-\rho(p_1)|^2 \  .
\eeq
\ed
Here $K$ is the \TR\ kernel (see \cite{EO07}), worth
\beq
K(p_1,p) = \frac12 \ \frac{\int_{p'=\sigma_a(p)}^{p} B(p_1,p')}{(\y(p)-\y(\sigma_a(p)))},
\eeq
where $\sigma_a( p)$ denotes the unique point such that $\rho(\sigma_a(p))=-\rho(p)$ in the disc around $a$.

Our hypothesis imply that
$B$ and $C$ are $<\infty$.

\subsection{The bounds}

The following theorem is the main result in this paper

\bt[Bound]\label{thbound}

If $2g-2+n>0$, $n\geq 1$ and  $p_1,\dots,p_n \in \curve_R$, we have the bound
\beq
\left|\frac{\omega_{g,n}(p_1,\dots,p_n)}{d\rho(p_1)\dots d\rho(p_n)}\right| \leq 
(n-1)! \ C_{g,n} \ \frac{C^{2g-2+n} \ B^{g-1+n}  }{ \left( \inf_{i\in\{1,\dots,n\}} |\rho(p_i)|\right)^{2d_{g,n}+2n} 
}
\eeq
where
\beq
d_{g,n}=3g-3+n
\quad , \quad
D_{g,n}=d_{g,n}+n
\eeq
and $C_{g,n}$ is the sequence defined by $C_{0,3}=1$, $C_{1,1}=1$, $C_{g,0}=0$, and by recursion
\bea
C_{g,n+1}
& = & \left( (n+1) C_{g-1,n+2} + \sum_{g_1+g_2=g, \ n_1+n_2=n}^{\text{stable}}
C_{g_1,n_1+1} \ C_{g_2,n_2+1} \right) \ \frac{(D_{g,n+1}+1)^{D_{g,n+1}+1}}{(D_{g,n+1})^{D_{g,n+1}}} \cr
& & + 2C_{g,n}  \ \frac{(2D_{g,n+1}+1)^{2D_{g,n+1}+1}}{3^3 (2D_{g,n+1}-2)^{2D_{g,n+1}-2}} \cr
\eea
where "stable" means $(g_i,n_i+1)\neq (0,1),(0,2)$.
\et

We shall use the following lemma, that we admit (proof straightforward)
\bl\label{lemmaeta}
If $k>0$ and $d>0$
\beq
\inf_{\eta\in ]0,1[} \frac{1}{(1-\eta)^k \eta^d} = \frac{(d+k)^{d+k}}{k^k \ d^d} \leq \frac{e^k}{k^k} \ (d+k)^k.
\eeq


\el

\textbf{Proof of theorem \ref{thbound}}.

Since this is the main result of this paper, we do the proof here in full detail.

First we write
\beq
W_{g,n}(p_1,\dots,p_n)=\frac{\omega_{g,n}(p_1,\dots,p_n)}{d\rho(p_1)\dots d\rho(p_n)} ,
\eeq
which is now a meromorphic function on $(\curve_R)^n$, with poles only at $\rho(p_i)=0$.

In all what follows we shall write
\beq
r_i=|\rho(p_i)|,
\eeq
\beq
r_{\min} = \min_i r_i,
\eeq
\beq
\eta_i = \frac{r_i}{r_{\min}} \geq 1.
\eeq

By definition of topological recursion \cite{EO07} we have
\bea\label{eqdefWgninproof}
 \omega_{g,n+1}(p_1,\dots,p_{n+1}) &= & \sum_{a\in \mathfrak R} \frac{1}{2\pi\ii} \oint_{p\in\mathcal C_a} K(p_1,p)   \Big[ \cr
 && \sum^{\text{stable}}_{g_1+g_2=g,I_1\sqcup I_2=\{p_2,\dots,p_{n+1}\}} \omega_{g_1,1+|I_1|}(p,I_1) \omega_{g_2,1+|I_2|}(\sigma_a(p),I_2)  \Big] \cr
&& + \sum_{a\in \mathfrak R} \frac{1}{2\pi\ii} \oint_{p\in\mathcal C_a} K(p_1,p)   \omega_{g-1,n+1}(p,\sigma_a(p),p_2,\dots,p_{n+1}) \cr
&& +  \sum_{j=2}^{n+1} \sum_{a\in \mathfrak R} \frac{2}{2\pi\ii} \oint_{p\in\mathcal C_a} \cr
&& \qquad K(p_1,p)   B(\sigma_a(p),p_j) \ \omega_{g,n}(p,p_2,\dots, \widehat{p_j} , \dots,p_{n+1})  \cr
\eea
where, for each term,  $\mathcal C_a$ is any small--enough circle around $a$, that we can choose to write as a circle in the coordinate $\rho(p)$ as:
\beq
\rho(p) = r \ e^{\ii \theta} \qquad , \qquad \theta\in [0,2\pi]  .
\eeq
"Small-enough" means that the value of the radius $r>0$ has to be chosen so that the circle doesn't enclose any point other than $a$, at which the integrand could have poles, in particular, since $K(p,p_1)$ has a pole at $\rho(p)=\pm \rho(p_1)$ , so we must have
\beq
r<r_1,
\eeq
and for the last line of \eqref{eqdefWgninproof}, for each value of $j$, since $B(\sigma_a(p),p_j)$ has a pole at $\rho(p)=-\rho(p_j)$, we must have 
\beq
r<r_j.
\eeq
We shall thus choose
\beq
r=\eta \ r_{\min} \quad , \quad \eta\in ]0,1[.
\eeq
The residue is independent of the value of $\eta\in ]0,1[$, and therefore we shall eventually choose the value of $\eta$ that will minimize the bound.

$\bullet$ We start with $(g,n)=(1,1)$:
\bea
 \omega_{1,1}(p_1) &= & \sum_{a\in \mathfrak R} \frac{1}{2\pi\ii} \oint_{p\in\mathcal C_a} K(p_1,p)   \Big[ B(p,\sigma_a(p)) \Big] 
\eea
From \eqref{defCKbound}, \eqref{defBBbound} we have for any $\eta\in ]0,1[$
\bea
\left| W_{1,1}(p_1) \right| 
&\leq  &  C B  \frac{1}{2\pi} \oint_{|\rho(p)|=r=\eta |\rho(p_1)| } \ \frac{|d\rho(p)/\rho(p)|}{|\rho(p_1)^2-\rho(p)^2| \ 4 \ |\rho(p)|^2}  \cr
&\leq  &  C B  \frac{1}{2\pi} \int_{0}^{2\pi}  \ \frac{d\theta}{4 \ (r_1^2-r^2) \ r^2}  \cr
&\leq  &  \frac{CB}{4 \ r_1^4} \ \frac{1}{(1-\eta^2)\eta^2}  \cr
&\leq  & \   \frac{CB}{r_1^4}  
\qquad \leftarrow \ \ \text{with} \ \eta=\frac{1}{\sqrt{2}},
\eea
so that the theorem holds with
\beq
C_{1,1} = 1.
\eeq

$\bullet$ Then for $(g,n)=(0,3)$, topological recursion gives:
\bea
 \omega_{0,3}(p_1,p_2,p_3) &= & 2\sum_{a\in \mathfrak R} \frac{1}{2\pi\ii} \oint_{p\in\mathcal C_a} K(p_1,p)   \Big[ B(p,p_2) \ B(\sigma_a(p),p_3) \Big] 
\eea
\bea
\left| W_{0,3}(p_1,p_2,p_3) \right| 
&\leq & 2CB^2 \frac{1}{2\pi} \oint_{|\rho(p)|=r} \frac{|d\rho(p)/\rho(p)|}{|\rho(p_1)^2-\rho(p)^2|} \   \frac{1}{|\rho(p)-\rho(p_2)|^2} \ \frac{1}{|\rho(p)+\rho(p_3)|^2} \cr
&\leq & 2CB^2 \frac{1}{2\pi} \int_{0}^{2\pi} \frac{d\theta}{(r_1^2-r^2)} \   \frac{1}{(r_2-r)^2 \ (r_3-r)^2} \cr
&\leq & \frac{2CB^2}{ r_{{\min}}^6} \  \frac{1}{(\eta_1^2-\eta^2) \ (\eta_2-\eta)^2 \ (\eta_3-\eta)^2} \cr
&\leq & \frac{2CB^2}{r_{{\min}}^6} \  \  \frac{1}{(1-\eta)^5} \cr
&\leq & \frac{2CB^2}{r_{{\min}}^6}  \qquad \leftarrow \ \ \text{with} \ \eta\to 0,
\eea
so that the theorem holds with
\beq
C_{0,3} = 1.
\eeq

$\bullet$ The bound shall then be proved by recursion.
Let $(g,n)$ such that $2g+n>2$.
Assume that the bounds are already proved for all $W_{g',n'}$ such that $2\leq 2g'+n'<2g+n+1$, we shall now prove it for $W_{g,n+1}$.

From the recursion hypotyhesis, and assuming that we choose the circle $\mathcal C_a$ of radius $r=\eta r_{\min}$, we have (we write $|I_1|=n_1$, $|I_2|=n_2$, so that $n_1+n_2=n$)
\bea
&& \left|\sum_{a\in \mathfrak R} \frac{1}{2\pi\ii} \oint_{p\in\mathcal C_a} d\rho(p)^2 K(p_1,p)     W_{g_1,1+n_1}(p,I_1) \ W_{g_2,1+n_2}(\sigma_a(p),I_2)  \right| \cr
&\leq &\frac{ C}{2\pi} \oint_{p\in\mathcal C_a} \frac{|d\rho(p)/\rho(p)|}{  |\rho(p_1)^2-\rho(p)^2|}    
\frac{n_1! \ C_{g_1,1+n_1} \ C^{2g_1-2+1+n_1} \ B^{g_1+n_1}}{|\rho(p)|^{2 d_{g_1,1+n_1}+2n_1+2} } \cr 
&& \frac{n_2 ! \ C_{g_2,1+n_2}\ C^{2g_2-2+1+n_2} \ B^{g_2+n_2}}{|\rho(p)|^{2 d_{g_2,1+n_2}+2n_2+2} }  \cr
&\leq & n_1! \ n_2 ! \ 
C_{g_1,1+n_1} \ C_{g_2,1+n_2} \ C^{2g-2+n+1} \ B^{g-1+n+1} \cr 
&& \frac{1}{2\pi} \int_{0}^{2\pi} \frac{d\theta}{ (r_1^2-r^2)}    \frac{1}{r^{2 d_{g_1,1+n_1}+2 d_{g_2,1+n_2}+2n+4}} \cr
&\leq & n_1! \ n_2 ! \  C_{g_1,1+n_1} \ C_{g_2,1+n_2} \ C^{2g-2+n+1} \ B^{g-1+n+1} \  \frac{1}{ (r_1^2-r^2)}    \frac{1}{r^{2 d_{g,n+1} +2n}} \cr
&\leq & \frac{n_1! \ n_2 ! \  C_{g_1,1+n_1} \ C_{g_2,1+n_2} \ C^{2g-2+n+1} \ B^{g-1+n+1} }{r_{{\min}}^{{2 d_{g,n+1} +2(n+1)}}} \ \frac{1}{ (\eta_1^2-\eta^2)  \ \eta^{2 d_{g,n+1} +2n}} \cr
&\leq & \frac{n_1! \ n_2 ! \  C_{g_1,1+n_1} \ C_{g_2,1+n_2} \ C^{2g-2+n+1} \ B^{g-1+n+1} }{r_{{\min}}^{{2 d_{g,n+1} +2(n+1)}}} \ \frac{1}{ (1-\eta^2)  \ \eta^{2 d_{g,n+1} +2n}} .
\eea
By a similar reasoning we get when $g>0$ 
\bea
&& \left|\sum_{a\in \mathfrak R} \frac{1}{2\pi\ii} \oint_{p\in\mathcal C_a} d\rho(p)^2 K(p_1,p)     W_{g-1,n+2}(p,\sigma_a(p),p_2,\dots,p_{n+1})  \right| \cr
&\leq &\frac{ C}{2\pi} \oint_{p\in\mathcal C_a} \frac{|d\rho(p)/\rho(p)|}{  |\rho(p_1)^2-\rho(p)^2|}    
\frac{(n+1)! \ C_{g-1,n+2} \ C^{2g-4+2+n} \ B^{g-1+n+1}}{|\rho(p)|^{2 d_{g-1,n+2}+2(n+2)} } \cr 
&\leq & \frac{(n+1)! C_{g-1,n+2} \ C^{2g-2+n+1} \ B^{g-1+n+1} }{r_{{\min}}^{{2 d_{g,n+1} +2(n+1)}}} \ \frac{1}{ (1-\eta^2)  \ \eta^{2 d_{g,n+1} +2n}} .
\eea

By a similar reasoning we get when $n>0$, and $j=2,\dots,n+1$:
\bea
&& \left|\sum_{a\in \mathfrak R} \frac{1}{2\pi\ii} \oint_{p\in\mathcal C_a} d\rho(p) K(p_1,p)  B(\sigma_a(p),p_j) \ W_{g,n}(p,p_2,\dots,\widehat{p_j},\dots,p_{n+1})  \right| \cr
&\leq & \frac{(n-1)! C_{g,n} \ C^{2g-2+n+1} \ B^{g-1+n+1} }{r_{{\min}}^{{2 d_{g,n+1} +2(n+1)}}} \ \frac{1}{ (\eta_1^2-\eta^2) \ (\eta_j-\eta)^2   \ \eta^{2 d_{g,n+1} +2n-2}} \cr
&\leq & \frac{(n-1)! C_{g,n} \ C^{2g-2+n+1} \ B^{g-1+n+1} }{r_{{\min}}^{{2 d_{g,n+1} +2(n+1)}}} \ \frac{1}{ (1-\eta)^3  \ \eta^{2 d_{g,n+1} +2n-2}} .
\eea
Using lemma \ref{lemmaeta},
the recursion hypothesis will be satisfied with
\bea
C_{g,n+1}
& = & \left( (n+1) C_{g-1,n+2} + \sum_{g_1+g_2=g, \ n_1+n_2=n}^{\text{stable}}
C_{g_1,n_1+1} \ C_{g_2,n_2+1} \right) \ \frac{(D_{g,n+1}+1)^{D_{g,n+1}+1}}{(D_{g,n+1})^{D_{g,n+1}}} \cr
& & + 2C_{g,n}  \ \frac{(2D_{g,n+1}+1)^{2D_{g,n+1}+1}}{3^3 (2D_{g,n+1}-2)^{2D_{g,n+1}-2}} .
\eea

\eproof

\br
The exponent of $1/r_{\min}$ i.e. $2d_{g,n}+2n$ is optimal, indeed it is reached for the Airy spectral curve, and is in agreement with \cite{eynclasses1,eynclasses2}.
\er

\br
But the coefficient $C_{g,n}$ is probably far from being optimal, it was obtained by bounding the integral by the integral of the absolute value, ignoring the phase oscillations, which could produce large cancellations. We are clearly overestimating here.
\er

\subsubsection{Factorial Bound }

\bt
We have the bounds:
\beq
C_{g,n} \leq t \  r^{-g} s^{-n}  \ (5g-5+3n) !  
\eeq
\beq
C_{g,n} \leq 9  \ (5g-5+3n) ! \  e^{4g-4+3n} 80^{2g-2+n} 3^{3-3g-3n} 14^{-g}
\eeq
where
\beq
s=\frac{27}{80} \ e^{-3},
\eeq
\beq
r=\frac{14\times 27}{80^2} \ e^{-4},
\eeq
\beq
t=\frac{3^5}{80^2} \ e^{-4}.
\eeq
\et
The bound can also be written
\beq
C_{g,n} \leq 9  \ (5g-5+3n) ! \  e^{4g-4+3n} 3^{5g-5+n}  14^{-g}.
\eeq

\proof

We shall prove the theorem by recursion. First observe that it is satisfied for $C_{0,3}=1$, $C_{1,1}=1$ and $C_{g,0}=0$.
Assume that it is satisfied for all $C_{g',n'}$ such that $2g'+n'<2g+n+1$. We shall now prove it for $C_{g,n+1}$.

Define
\beq
A_{g,n} = 5g-5+3n 
\quad , \quad \kappa_{g,n}=2g-2+n
\quad , \quad  D_{g,n}=3g-3+2n.
\eeq
For stable $(g,n)$ (i.e. $(g,n)\neq (0,1),(0,2)$) and with $n\geq 1$ we have
\beq
\kappa_{g,n}\geq 1 \quad , \quad A_{g,n}\geq 3\quad , \quad D_{g,n}\geq 2.
\eeq
We shall need the following inequalities:
\begin{itemize}
\item
\beq
D_{g,n}+1 \leq   D_{g,n}+\kappa_{g,n} =A_{g,n} .
\eeq
\item
\beq
n=2A_{g,n}-5\kappa_{g,n} \leq 2A_{g,n}-5=2(A_{g,n}-1)-1
\eeq
\item
for all $u\in ]0,\frac52[$ we have
\beq
g-1 = \frac{1}{5-2u}(A_{g,n}-u\kappa_{g,n}-3n+un) \leq \frac{A_{g,n}-3}{5-2u}
\quad\implies \quad g+1  \leq \frac{A_{g,n}+7-4u}{5-2u}.
\eeq
The case $u=\frac94$ gives
\beq
g+1  \leq 2(A_{g,n}-2).
\eeq

\item The number of stable pairs $(g_1,1+n_1),(g_2,1+n_2)$ such that $g_1+g_2=g$ and $n_1+n_2=n$, is:
\beq
(g+1)(n+1)-4 \leq 4(A_{g,n+1}-2)(A_{g,n+1}-1-\frac12).
\eeq

\item
We have
\beq
A_{g-1,n+2} = A_{g,n+1} -2
\eeq
\beq
A_{g_1,n_1+1}+A_{g_2,n_2+1}-1 = A_{g,n+1} - 3
\eeq
\beq
A_{g,n} = A_{g,n+1} -3.
\eeq

We shall use the property that for any $a,b$ strictly positive integers, we have
$a! b! \leq (a+b-1)!$.
This implies that
\bea
A_{g_1,n_1+1}! A_{g_2,n_2+1}! 
\leq (A_{g_1,n_1+1}+A_{g_2,n_2+1}-1)! 
= (A_{g,n+1}-3)!
\eea
\beq
A_{g-1,n+2}! = (A_{g,n+1}-2)! = (A_{g,n+1}-3)! (A_{g,n+1}-2)
\eeq
\beq
A_{g,n}! = (A_{g,n+1}-3)! 
\eeq

\end{itemize}

From lemma \ref{lemmaeta}, we have:
\bea
C_{g,n+1}
& \leq  & \left(  (n+1) C_{g-1,n+2} + \sum_{g_1+g_2=g, \ n_1+n_2=n}^{\text{stable}}
C_{g_1,n_1+1} \ C_{g_2,n_2+1} \right) \ e (D_{g,n+1}+1) \cr
& & + 2C_{g,n}  \  \frac{e^3}{3^3} (2D_{g,n+1}+1)^3 ,
\eea
now using the recursion hypothesis we have
\bea
C_{g,n+1}
& \leq  & t \ r^{-g} s^{-n-1} \Big( \frac{r}{s} (n+1) A_{g-1,n+2}! \cr
&& + t s^{-1} \sum_{g_1+g_2=g, \ n_1+n_2=n}^{\text{stable}}
A_{g_1,n_1+1}! \ A_{g_2,n_2+1}! \Big) \ e (D_{g,n+1}+1) \cr
& & +  t \ r^{-g} s^{-n-1} A_{g,n}!  \  \frac{2^4 e^3 s} {3^3 } (D_{g,n+1}+\frac12)^3 ,
\eea
and thus
\bea
\frac{C_{g,n+1}}{t \ r^{-g} s^{-n-1} A_{g,n+1}!}
& \leq  & \frac{1}{A_{g,n+1}(A_{g,n+1}-1)(A_{g,n+1}-2)} \Big( \Big(\frac{r}{s} (n+1) (A_{g,n+1}-2) \cr
&& + \frac{t}{s} ((g+1)(n+1)-4) \Big) \ e (D_{g,n+1}+1) \cr
& & +    \frac{2^4 e^3 s}{3^3 } (D_{g,n+1}+\frac12)^3 \Big),
\eea
Remark that $D_{g,n+1}+1\leq A_{g,n+1}$ and $D_{g,n+1}+\frac12\leq A_{g,n+1}$, therefore
\bea
\frac{C_{g,n+1}}{t \ r^{-g} s^{-n-1} A_{g,n+1}!}
& \leq  & \frac{e}{(A_{g,n+1}-1)(A_{g,n+1}-2)} \Big( \Big(\frac{r}{s} (n+1) (A_{g,n+1}-2) \cr
&& + \frac{t}{s} ((g+1)(n+1)-4) \Big)  \cr
& & +    \frac{2^4 e^2 s}{3^3 } (D_{g,n+1}+\frac12)^2 \Big),
\eea

We define
\beq
er/s=c''= 14/80
\eeq
\beq
et/s=c=9/80
\eeq
\beq
2^4 e^3 s/3^3  = c'=16/80.
\eeq

writing $A=A_{g,n+1}$, we have
\bea
&& c'' (n+1) (A_{g,n+1}-2)+c((g+1)(n+1)-4)+c'(D_{g,n+1}+\frac12)^2 \cr
& \leq & c''(2(A-1)-1)(A-2) + c (2(A-2)(2(A-1)-1)-4)+c'(A-\frac12)^2 \cr
& \leq & (2c''+4c)(A-1)(A-2) -  (c''+2c)(A-2) -4c+c'(A^2-A+\frac14) \cr
& \leq & (2c''+4c)(A-1)(A-2) -  (c''+2c)(A-2) -4c+c'((A-1)(A-2) +2A -2+\frac14) \cr
& \leq & (2c''+4c+c')(A-1)(A-2) -  (c''+2c)(A-2) -4c+c'(2A -2+\frac14) \cr
& \leq & (2c''+4c+c')(A-1)(A-2) -  (c''+2c-2c')(A-2) -4c+c'(2+\frac14) \cr
\eea
We have
\beq
c''+2c-2c'= 0
\eeq
\beq
4c-\frac94 c'= 0
\eeq
and
\beq
2c''+4c+c'=1.
\eeq
This implies
\beq
c'' (n+1) (A_{g,n+1}-2)+c((g+1)(n+1)-4)+c'(D_{g,n+1}+\frac12)^2  \leq  (A_{g,n+1}-1)(A_{g,n+1}-2) 
\eeq
which implies the bound for $C_{g,n+1}$.
\eproof

\subsection{Bounds for $F_g$}

For $g\geq 2$ we have \cite{EO07}
\beq
F_g = \frac{1}{2g-2} \sum_{a\in \mathfrak R} \frac{1}{2\pi\ii} \oint_{\mathcal C_a} \omega_{g,1}(p) \Phi(p)
\eeq
where $d\Phi=(y-y(a)) dx$.
Our assumption that $y$ behaves like a square-root implies that $\Phi(p)-\Phi(a)$ behaves like $O(\rho(p)^3)$.
Let us define
\beq
\td C = \frac{1}{\# \mathfrak R}  BC \sup_{p\in \curve_R} |\Phi(p)-\Phi(a)|\  |\rho(p)|^{-3}.
\eeq

\bt
For $g\geq 2$ we have
\beq
|F_g| \leq \td C  C^{2g-2} B^{g-1} \ \frac{1}{R^{6g-6}} \ \frac{C_{g,1}}{2g-2}.
\eeq

\beq
|F_g| \leq \td C \frac{9}{80 e} C^{2g-2} B^{g-1} \ \frac{1}{R^{6g-6}} \ r^{-g} \frac{(5g-2)!}{2g-2}.
\eeq

\et

\proof
Choosing the circle of radius $|\rho(p)|=R$, one has
\bea
(2g-2)|F_g|
&\leq & \frac{1}{2\pi} \frac{\td C}{BC} \int_{0}^{2\pi} C^{2g-2+1} B^{g-1+1} C_{g,1} \frac{R^3 }{R^{2d_{g,1}+2}} R d\theta \cr
&\leq & \td C C^{2g-2} B^{g-1} C_{g,1} \frac{1}{R^{2d_{g,1}-2}}  \cr 
&\leq & \td C C^{2g-2} B^{g-1} C_{g,1} \frac{1}{R^{6g-6}} .
\eea

\eproof

Remark that $R$ was constrained by the condition that  discs $|\rho(p)|<R$ are disjoints, in other words $R$ somehow measures the "distance between ramification points", and thus we recover the well known fact that $F_g$ diverges when ramification points meet.

\section*{Conclusion:  Borel transform and resurgence}

In this article, we have showed that, under reasonable generic asumptions $F_g(\spcurve) $ has a factorial growth at large $g$, of speed at most $(5g)!$.
We already pointed out that this is an upper bound, probably overestimated, and indeed for most known examples, $F_g$ has actually a factorial growth of order $(2g)!$.

Let us assume that $F_g$ has a factorial growth of order $(\beta g)!$ with $\beta\leq 5$.

We may define
\beq
\hat F(\spcurve,s) = \sum_{g=0}^\infty \frac{s^{\beta g}}{(\beta g)!} \ F_g(\spcurve)
\eeq
which is absolutely convergent in a disc.

It may happen that it is an entire function convergent in the whole complex plane $\CC$ (this is the case where the growth of $F_g$ was actually slower than $(\beta g)!$, and one could choose a smaller value of $\beta$).

If $\hat F(\spcurve,s)$ would be analytically continuable beyond its convergence disc, up to $\infty$, we would recover $F$ by the Laplace transform
\beq
F(\spcurve,\hbar) = \hbar^{-2-\frac2\beta} \int_0^\infty ds e^{-s \hbar^{-\frac2\beta}} \hat F(\spcurve,s).
\eeq
This requires to know if $\hat F(\spcurve,s)$ can be analytically continued beyond its convergence disc, up to $\infty$, in other words this requires to know if $F_g$ is a resurgent series \cite{resurgence}.

Equivalently this needs to know where the singularities of $\hat F(\spcurve,s)$ can be, or what are the possible divergences at $\infty$.

If $\hat F$ has singularities at finite distance, we may get contributions to $F$ of the type
\beq
e^{-s_{\text{sing}} \hbar^{-\frac2\beta}}.
\eeq
If $\beta=2$ we would get corrections in $e^{-\hbar^{-1}}$.

If $\beta>2$ and $\hat F$ is an entire function and behaves at $\infty$ as
\beq
\hat F\sim e^{s^\alpha}
\eeq
We may get contributions to $F$ of the type
\beq
e^{-\hbar^{\frac{-2\alpha}{\beta(\alpha-1)}}}.
\eeq
For instance if $\alpha=\frac\beta{\beta-2}$ we would get corrections in $e^{-\hbar^{-1}}$.

We shall study the resurgence properties in a forthcoming work...

\section*{Acknowledgments}

This work was supported by the ERC Starting Grant no. 335739 ``Quantum fields and knot homologies'' funded by the European Research Council under the European Union's Seventh Framework Programme. 
It is also partly supported by the ANR grant Quantact : ANR-16-CE40-0017.
I wish to thank IHES and M. Kontsevich.

\end{document}